# A Grid Service Broker for Scheduling Distributed Data-Oriented Applications on Global Grids

Srikumar Venugopal[1], Rajkumar Buyya[1] and Lyle Winton[2]

[1]Grid Computing and Distributed Systems Laboratory
Dept. of Computer Science and Software Engineering
The University of Melbourne, Australia
{srikumar,raj}@cs.mu.oz.au

[2]Experimental Particle Physics Group
School of Physics
The University of Melbourne, Australia
winton@ph.unimelb.edu.au

**Abstract:**

*The next generation of scientific experiments and studies, popularly called as e-Science, is carried out by large collaborations of researchers distributed around the world engaged in analysis of huge collections of data generated by scientific instruments. Grid computing has emerged as an enabler for e-Science as it permits the creation of virtual organizations that bring together communities with common objectives. Within a community, data collections are stored or replicated on distributed resources to enhance storage capability or efficiency of access. In such an environment, scientists need to have the ability to carry out their studies by transparently accessing distributed data and computational resources. In this paper, we propose and develop a Grid broker that mediates access to distributed resources by (a) discovering suitable data sources for a given analysis scenario, (b) suitable computational resources, (c) optimally mapping analysis jobs to resources, (d) deploying and monitoring job execution on selected resources, (e) accessing data from local or remote data source during job execution and (f) collating and presenting results. The broker supports a declarative and dynamic parametric programming model for creating grid applications. We have used this model in grid-enabling a high energy physics analysis application (Belle Analysis Software Framework). The broker has been used in deploying Belle experiment data analysis jobs on a grid testbed, called Belle Analysis Data Grid, having resources distributed across Australia interconnected through GrangeNet.*

## 1  Introduction

The next generation of scientific experiments and studies, popularly called as e-Science [15], is carried out by communities of researchers from different organizations that span national and international boundaries. These experiments involve geographically distributed and heterogeneous resources such as computational resources, scientific instruments, databases and applications. The data in these experiments is usually massive and distributed across numerous institutions for various reasons including, the inherent distribution of data sources; large-scale storage and computational requirements; to ensure high-availability and fault-tolerance of data; and caching to provide faster access. Some well-known scientific experiments of this nature include the CERN-led ATLAS [13] and CMS [14] experiments and KEK's Belle [16] experiment. The users in such complex environments should able to carry out analysis of the data generated by the experiments by transparently accessing distributed datasets and computational resources. They should also be able to share the results of their analysis with the rest of the community.

Grid computing [1] enables the creation of virtual organizations [17] by bringing together communities with common objectives. Grid platforms support sharing, exchange, discovery, selection, and aggregation of geographically/Internet-wide distributed heterogeneous resources — such as computers, databases, visualization devices, and scientific instruments. Recently, Data Grids [2] have evolved to tackle the twin challenges of large datasets and multiple data repositories at distributed locations in data-intensive computing environments [3]. However, the harnessing of the complete power of grids remains to be a challenging problem for users due to the complexity involved in the creation and composition of applications and their deployment on distributed resources.

Resource brokers hide the complexity of grids by transforming user requirements into a set of jobs that are scheduled on the appropriate resources, managing them and collecting results when they are finished. A resource broker in a data grid must have the capability to locate and retrieve the required data from multiple data sources and to redirect the output to storage where it can be retrieved by processes downstream. It must also have the ability to select the best data repositories from multiple sites based on availability of files and quality of data transfer. In this paper, one such broker called the Gridbus Broker providing services relevant to data-intensive environments is presented. Its application to the high-energy physics domain is discussed by illustrating its use within the Belle Analysis Data Grid and the results of experiments that have been conducted on it are presented.

## 2   Related Work

In the context of Data Grid, it is worthwhile to mention the Storage Resource Broker (SRB) [4] from San Diego Supercomputing Centre (SDSC) which provides middleware for storing datasets over a network and accessing them. However, it does not deal with application execution directly. Hence, it is similar to other data replication mechanisms such as Grid Data Mirroring Package (GDMP) [5] and Giggle [6]. The European DataGrid [8] has its own resource broker which is installed in a central machine that receives requests and then decides to dispatch jobs according to system parameters. Cactus [7] is a numerical problem solving environment for scientists which supports Data Grid features through the use of MPICH-G and Globus. However, applications in Cactus environment have to be written in MPI which implies that a legacy application cannot be adapted as such to be run on a grid. NILE [30] is a grid-computing environment for high-energy physics constructed using CORBA but is limited to that domain.

The Gridbus broker extends the Nimrod-G [10] computational Grid resource broker model to distributed data-oriented grids. Nimrod-G specializes in parameter-sweep computation. However, the scheduling approach within Nimrod-G aims at optimizing user-supplied parameters such as deadline and budget [11] for computational jobs only. It has no functions for accessing remote data repositories and for optimizing on data transfer. The Gridbus broker also extends Nimrod-G's parametric modeling language by supporting dynamic parameters, i.e. parameters whose values are determined at runtime.

Like Nimrod-G, the AppLeS PST [9] supports deployment of parameter-sweep applications on computational grids, but its adaptive scheduling algorithm emphasizes on data-reuse. The users can identify common data files required by all jobs and the scheduling algorithm replicates these data files from the user node to computational nodes. It tries to re-use the replicated data to minimize the data transmission when multiple jobs are assigned to the same resource. However, multiple repositories of data are not considered within this system and therefore, this scheduling algorithm is not applicable to Data Grids.

Ranganathan and Foster [12] have conducted simulation studies for various scheduling scenarios within a data grid. Their work recommends decoupling of data replication from computation while scheduling jobs on the Grid. It concludes that it is best to schedule jobs to computational resources that are closest to the data required for that job, but the scheduling and simulation studies are restricted to homogeneous computational nodes with a simplified First-In-First-Out (FIFO) strategy within local schedulers.

Similar to [12], our work focuses on a resource scheduling strategy within a Data Grid but we concentrate on adaptive scheduling algorithms and brokering for heterogeneous resources that are shared by multiple user jobs. In addition, the scheduling strategy has been implemented within the Gridbus broker and its feasibility to support the deployment of distributed data-intensive applications (e.g. KEK Belle high-energy physics experiment data analysis) within a real Grid testbed (e.g., Australian Belle Analysis Data Grid) has been evaluated.

## 3   Architecture

### 3.1   Data Grid Overview and Brokering

A data-intensive computing environment can be perceived as a real-world economic system wherein there are producers and consumers of data. Producers are entities which generate the data and control its distribution via mirroring at various replica locations around the globe. They lay down policies for replication that are guided by various criteria such as minimum bandwidth, storage and computational requirements, data security and access restrictions and data locality issues. However, information about the data replicas is assumed to be available through a data catalogue mechanism such as the Globus Replica Catalog [18]. An example of such a system would be the tier-level model proposed by the MONARC [19] group within CERN for replicating the data produced by the Large Hadron Collider (LHC) [20] for use within the ATLAS and CMS collaborations. The consumers in this system would be the users or, by proxy, their applications which need to analyse this data to produce meaningful results. The users may want to investigate specific datasets out of a set of hundreds and thousands and may have specific application requirements that need not be fulfilled at every computational site. A sample scenario for such a data-intensive computing environment and the role of the broker in it is illustrated in Figure 1.

The idea of a computational economy [11] helps in creating a service-oriented computing architecture where service providers offer paid services associated with a particular application and users, based on their requirements, would optimize by selecting the services they require and can afford within their budget. In a non-profit scientific collaboration, this notion can be used to regulate the usage of precious resources such as network bandwidth, computational and storage resources by providing users with tokens that can be redeemed against resource usage. To realize this scenario, service publication, metering and accounting mechanisms such as the Grid Market Directory (GMD) [21], the Grid Trade Server (GTS) [22] and GridBank [23] respectively, have been developed. Though details of the integration of economic factors within the architecture are beyond the scope of this work, they are presented here as these mechanisms are integral to a data-intensive environment.

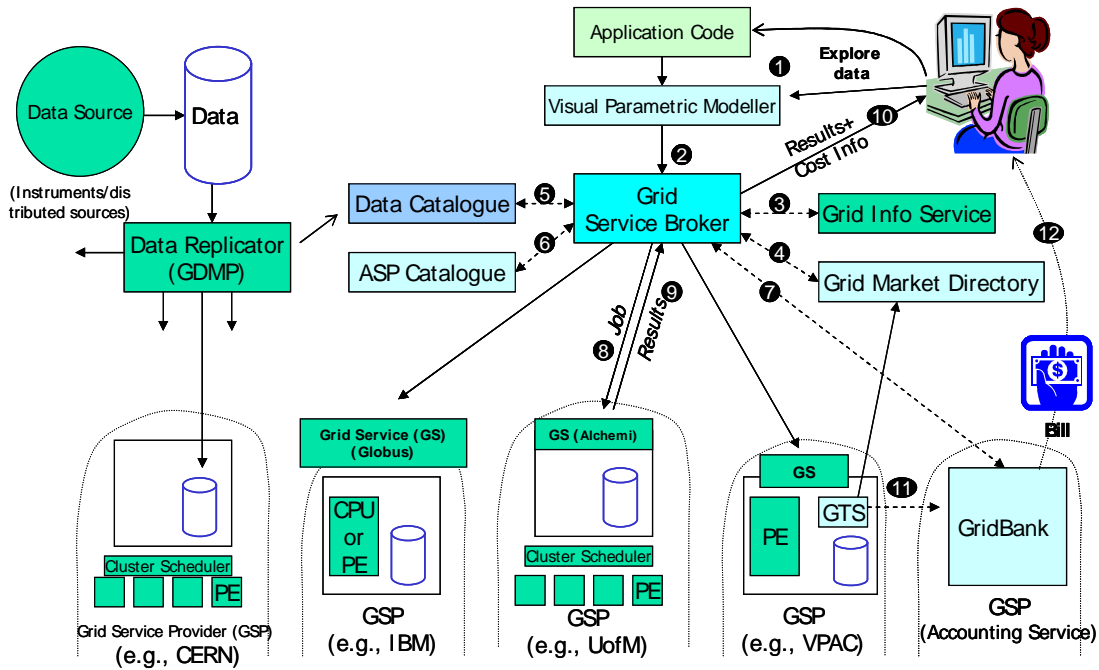

**Figure 1: A sample Data Grid and Analysis scenario.**

In the Data Grid environment depicted in Figure 1, the steps involved in analysing distributed data are as follows. The application code is the legacy application that has to be executed on a grid. The users compose their application as a distributed application (e.g., parameter sweep) using visual application development tools (Step 1). The parameter-sweep model of creating several independent jobs is well suited for grid computing environments wherein challenges such as load volatility, high network latencies and high probability of failure of individual nodes make it difficult to adopt a programming approach which favours tightly coupled systems. Accordingly, this has been termed as a "killer application" for the Grid [10]. Visual tools allow rapid composition of applications for grids while taking away the associated complexity.

The user's analysis and quality-of-service requirements are submitted to the Grid resource broker (Step 2). The Grid resource broker performs resource discovery based on user-defined characteristics, including price, using the Grid information service and the Grid Market Directory (Steps 3&4). The broker identifies the list of data sources or replicas and selects the optimal ones (Step 5). The broker also identifies the list of computational resources that provides the required application services using the Application Service Provider (ASP) catalogue (Step 6). The broker ensures that the user has the necessary credit or authorized share to utilise resources (Step 7). The broker scheduler maps and deploys data analysis jobs on resources that meet user quality-of-service requirements (Step 8). The broker agent on a resource executes the job and returns results (Step 9). The broker collects the results and passes them to the user (Step 10).The metering system charges the user by passing the resource usage information to the accounting system (Step 11). The accounting system reports resource share allocation or credit utilisation to the user (Step 12).

## 3.2 Gridbus Data Grid Service Broker

The architecture of the Gridbus broker is shown in Figure 2. The inputs to the broker are the tasks and the associated parameters with their values. These can be specified within a "plan" file that specifies the tasks and the types of the parameters and their values for these tasks.

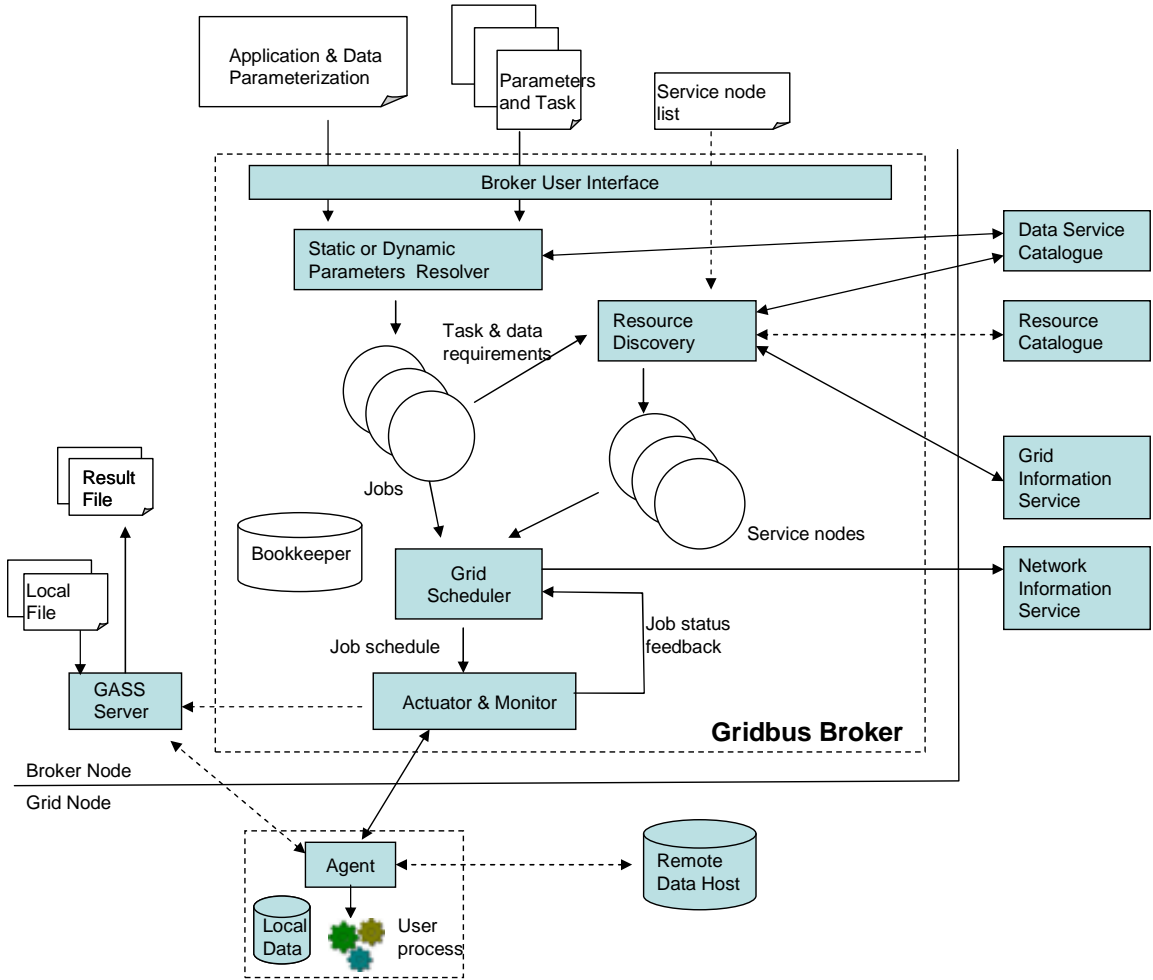

**Figure 2: Gridbus broker architecture and its interaction with other Grid entities.**

A task is a sequence of commands that describe the user's requirements. For example, the user may specify an application to be executed at the remote site. But, the requirements may also require an input file to be copied over before execution and the results to be returned back. Hence, a task encapsulates this information within its description. A task is accompanied by parameters which can either be static or dynamic. A static parameter is a variable whose domain is well-defined either as a range of values, as a single static value or as one among a set of values. A dynamic parameter has either an undefined or an unbounded domain whose definition or boundary conditions respectively, have to be established at runtime. As an example, in the current implementation, a parameter type has been defined which describes a set of files over which the application has to be executed. This set can be described as a wildcard search within a physical or a logical directory, to be resolved at runtime, thus creating a dynamic parameter.

The task requirements drive the discovery of resources such as computational nodes and data resources. The resource discovery module gathers information from remote information services such as the GMD or Grid Index Information Service (GIIS) [25] for availability of compute resources. Optionally, the list of available compute resources can be provided by the user to the broker. The broker also interacts with the information service on each computational node to obtain its properties. Data files can be organised as Logical File Names (LFNs) within a virtual directory structure using a Replica/Data Service Catalog. Each LFN maps to one or many Physical File Names (PFNs) located somewhere on Grid, usually specified via URLs. The virtual directory structure is organised into catalogues and further into collections. As mentioned before, the LFNs to be processed may be typically specified by a dynamic parameter and the broker will resolve this to the appropriate physical file location(s).

The task description, i.e. the task along with its associated parameters, is resolved or "decomposed" into jobs. A job is an instantiation of the task with a unique combination of parameter values. It is also the unit of work that is sent to a Grid node. The set of jobs along with the set of service nodes are an input to the scheduler. The scheduler matches the job requirements with the services and dispatches jobs to the remote node. For jobs requiring remote data, it interacts with a network monitoring service to obtain the information about current available bandwidth between the data sources and the compute resources. It then uses this information to schedule jobs by optimizing on the amount of data transfer involved. In the current implementation, the Network Weather Service (NWS) [33] has been used to obtain this information. The scheduling algorithm is described in more detail in the next section.

The jobs are dispatched to the remote node through the Actuator component. The Actuator submits the job to the remote node using the functionality provided by the middleware running on it. The Actuator has been designed to operate with different Grid middleware frameworks and toolkits such as Globus 2.4 [34] that primarily runs on Unix-class machines and Alchemi [36], which is a .NET based grid computing platform for Microsoft Windows-enabled computers. Hence, it is possible to create a cross-platform grid implementation using the Gridbus broker. The task commands are encapsulated within an Agent which is dispatched to and executed on the remote machine. If a data file has been associated with the job and a suitable data host identified for that file, then the Agent obtains the file through a remote data transfer from the data host. Additionally, it may require some configuration or input parameter files that it obtains from the broker through a mechanism such as a GASS [26] (Globus Access to Secondary Storage) Server. These files are assumed to be small and in tens or hundreds of kilobytes which impact the overall execution time of a job negligibly whereas the data files are in the range of megabytes or larger. On completion of execution, the Agent returns any results to the broker and provides debugging information. The Monitoring component keeps track of job status – whether the jobs are queued, executing, finished successfully or failed. It updates the status of the jobs which is fed back to the scheduler to update its estimates of the rate of execution and of the performance of the compute resources. The Bookkeeper keeps a persistent record of job and resource states throughout the entire execution.

### 3.3 Scheduling Algorithm

The scheduler within the broker looks at a data grid from the point of view of the data. It perceives a data-intensive computing environment as a collection on data hosts, or resources hosting the data, surrounded by compute servers, i.e. resources offering the computation service. The "network proximity" of a compute resource to a data host is a measure of the available bandwidth between the resources. Some of the data resources may have computation facilities too, in which case there is assumed to be nearly infinite bandwidth between the data host and the compute resource at the same site. The algorithm for the scheduler is listed in Figure 3.

The scheduler minimizes the amount of data transfer involved while executing a job by dispatching jobs to compute servers which are close to the source of data. A naïve way of achieving this would be to run the jobs only on those machines that contain their data. But, the data hosts may not have the best computational resources. Hence, the scheduler selects an appropriate compute resource to execute a job based on factors such as capability and performance of the resource, bandwidth available from the compute resource to the data host that contains the data file required for the job and the cost of data transfer. A detailed experimental analysis and performance evaluation of this scheduling algorithm for different scenarios is presented in Section 4 with a case study on the deployment of high energy physics application within a Data Grid environment.

> **Initialisation**
> 1. Identify characteristics, configuration, capability, and suitability of compute resources using the Grid information services (GIS).
> 2. From the task definition, obtain the data query parameters (if present), such as the logical file name
>    a. Resolve the data query parameter to obtain the list of Logical Data Files (LDFs) from the Data Catalog
>    b. For each LDF, get the data sources or Data Hosts that store that file by querying the Data Catalog.
>
> **Scheduling Loop**
> Repeat while there exists *unprocessed jobs.* [This step is triggered for each scheduling event. The event period is a function of job processing time, rescheduling overhead, resource share variation, etc.]:
> 3. For each compute resource, predict and establish the *job consumption rate* or the *available resource share* through the measure and extrapolation strategy taking into account the time taken to process previous jobs. Use this estimate along with its current commitment to determine expected job completion time.
> 4. If any of the compute resource has jobs that are yet to be dispatched for execution and there is variation in resource availability in the Grid, then move such jobs to the Unassigned-Jobs-List.
> 5. Repeat until all unassigned jobs are scheduled or all compute resources have reached their maximum job limit.
>    a. Select the next job from the Unassigned-Jobs-List.
>    b. Identify all Data Hosts that contain the LDF associated with the job.
>    c. Create a Data-ComputeResource-List for the selected job:
>       - For each data host, identify a compute resource that can complete the job earliest given its current commitment, job completion rate, and data transfer time using current available bandwidth estimates.
>    d. Select a data host and compute resource pair with the earliest job completion time from the Data-ComputeResource-List.
>    e. If there exists such a resource pair, then assign the job to the compute resource and remove it from the Unassigned-Jobs-List.
> 6. End of scheduling loop.

**Figure 3: Adaptive scheduling algorithm for Data Grid.**

## 3.4  Design and Implementation

The Gridbus broker has been implemented in Java so that it can be deployed in Web-enabled environments such as Tomcat-driven portals and also be used from the command line. It interfaces to nodes running Globus[34] using the Java Commodity Grid (CoG) Kit [27] and to Alchemi nodes using the Alchemi Cross-Platform Manager Interface described in [36]. A UML (Unified Modelling Language) class diagram that displays the core entities within the broker and their associations is shown in Figure 4.

The main design entities within the broker are:

1. Compute Server: The ComputeServer object describes a node on the grid. It holds the properties for that node e.g. what middleware it is running, its architecture and its operating system. It also monitors the rate of progress by keeping track of the number of jobs that are done, failed or executing on the corresponding remote server.

   The ComputeServer has been extended for different middlewares like Globus and Alchemi. However, the scheduler has a platform-independent view of the nodes and is concerned only with their performance. Therefore, it is possible for the broker to operate across different middlewares.

2. Job: - A job is an abstraction for a unit of work assigned to a node. A Job has the following structure:
   a. Variables: - A variable holds the designated parameter value for a job. A variable can hold a range of values or a set of values or a single value in which case it is called a single variable
   b. Task: - As described above, a task is the description of what has to be done by the job. A task is composed of a set of commands. There are three types of commands
      i. Copy Command: Instructs the broker to copy a file from the source to the destination. It can be used for either copying the files from the broker host to the remote node or vice versa. At present, there is no provision to copy from one node to another node without the broker intervening in between. A special case of copy command is the Multiple Copy (MCopy)

command which instructs the broker to copy multiple files as described with a wildcard (*, ?, etc.)
   ii. Execute Command: Instructs the broker to execute the application given as a parameter to this command on the remote node.
   iii. Substitute Command: Instructs the broker to substitute the values whenever it encounters a variable name within a text file. This is particularly useful when a configuration file has to be modified for each job.

When a job is submitted to a server, the control is passed to the JobWrapper associated with the job. This object takes charge of translating the job instructions to be understood by the middleware that is running on the designated ComputeServer. After a successful submission, the JobWrapper starts up the JobMonitor which tracks the job and updates its status. Otherwise, it throws an exception and quits.

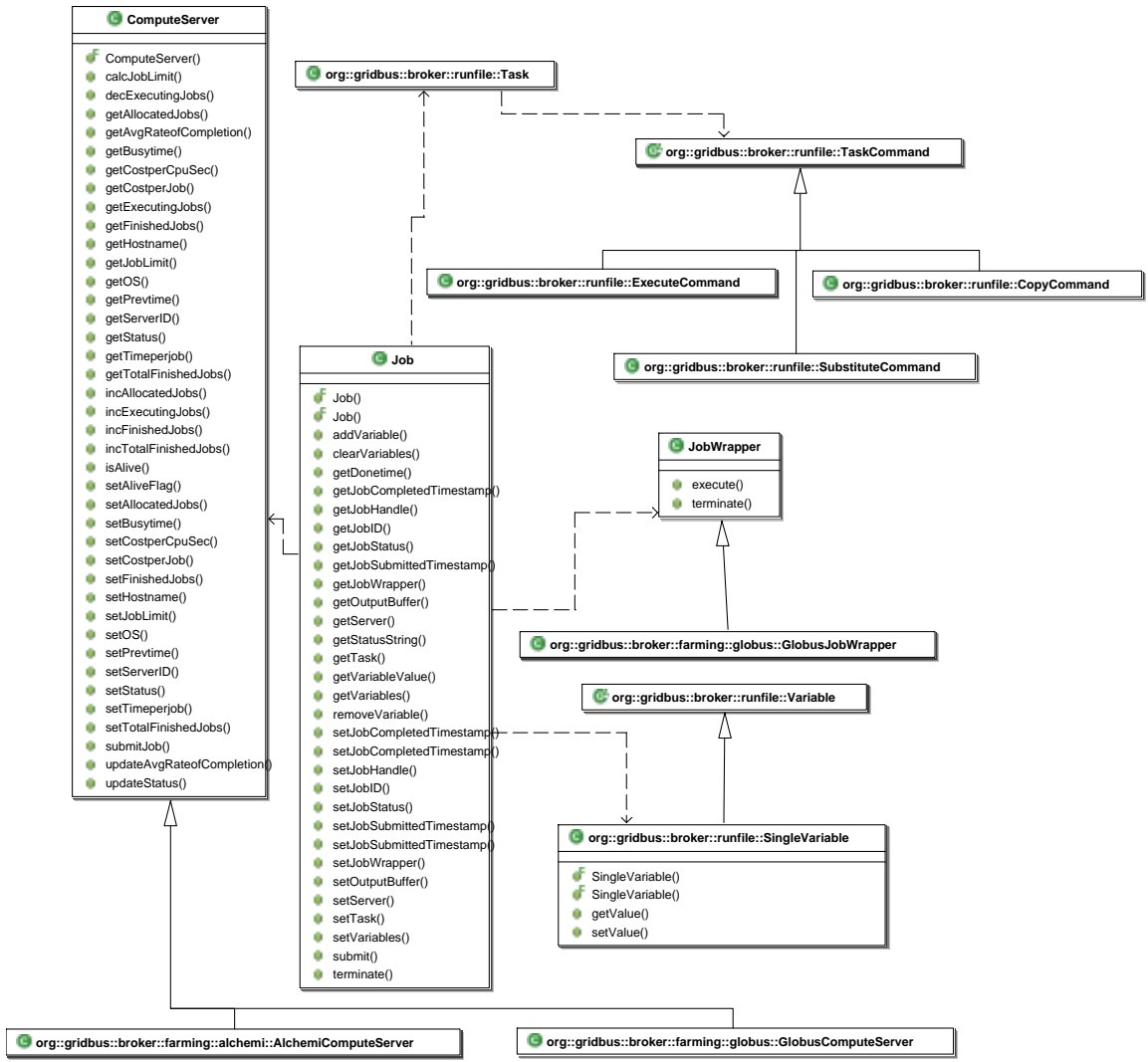

**Figure 4: UML Diagram of core broker components**.

3. Data Hosts: - Data Hosts are nodes on which data files have been stored. These objects store the details of the data files that are stored on them such as their path on the disk and the protocol used to access them. The Data Host objects also maintain a list of the compute resources sorted in the descending order of available bandwidth from the host.
4. Data Files: - The Data File object stores attributes of input files that are required for an application

such as size and location. A Data File object links to the different Data Hosts that store that file.

Other than these, the broker also has a Farming Engine component which is the central component of the broker and its point of contact with other applications. The farming engine initializes the objects within the broker and invokes the scheduler to distribute the jobs. User Interfaces such as command line or web interfaces interact with the farming engine to supply inputs such as parameterized plan files and list of compute resources and to obtain data about the progress of execution.

## 4  A Case Study in High Energy Physics

High Energy Physics (HEP) is a fundamental science studying matter at the very smallest scales. Probing this frontier requires accelerators of great complexity, typically beyond the means of any single country. Equally, since experiments in HEP are large and technically sophisticated, they necessarily involve international collaboration between many institutes over very long time scales.

Computing resource requirements for HEP are increasing exponentially because of advancements in particle accelerators and increasing size of collaborations. Modern experiments will have to provide access to petabytes of data, hundreds of teraflops of computing power, for thousands of researchers located in many institutions around the world. Existing techniques for analysis using high performance computing will not be sufficient. The CERN LHC particle accelerator is a case in point of how current computational facilities will prove inadequate for the next generation of scientific experiments and is frequently cited as a justification for the need for data grids in experimental high energy physics [28].

### 4.1  The Belle Project

Charge-Parity (CP) violation was first observed in 1964, by studying the decays of K-mesons. Briefly C is the symmetry operation of particle - antiparticle inversion, and P that of space inversion. The issue today is whether the Standard Model(SM) of Physics offers a complete description of CP violation, or, more importantly, whether new physics is needed to explain it. Answering this question requires very detailed study of this subtle effect.

A new and exciting approach to exploring CP violation stems from the prediction that its effects should be observed to be considerably larger in the decays of B-mesons, particles containing the b (or bottom) quark [31]. The Belle experiment, built and operated by a collaboration of 400 researchers across 50 institutes from 10 countries, is probing CP-violation by studying the decay of the B-mesons produced in the KEKB accelerator at the Japanese High Energy Accelerator Research Organization (KEK) in Tsukuba. The increasing efficiencies of the KEKB accelerator have led to an increase in the rate of data production from the Belle experiment. The current experiment and simulation data set is tens of terabytes in size. While this increase is extremely desirable for the study of B-meson decays, it begins to pose problems for the processing and access of data at geographically remote institutions, such as those within Australia. Hence, it is important for Data Grid techniques to be applied in this experiment [29].

### 4.2  The Testbed

The Belle Analysis Data Grid testbed has been set up in Australia in collaboration with IBM is shown in Figure 5. The testbed resources are located in Sydney (Dept. of Physics, University of Sydney), Canberra (Australian National University), Melbourne (School of Physics and the Dept. of Computer Science, University of Melbourne) and Adelaide (Dept. of Computer Science, University of Adelaide). All the nodes in the testbed, except for the one in Adelaide, are connected via GrangeNet (Grid And Next Generation Network) [34]. GrangeNet is a three year program to install, develop and operate a multi-gigabit network supporting grid and advanced communications services across Australia. Hence, there is a higher bandwidth between the Melbourne, Canberra and Sydney resources. The machines in this testbed, their capabilities and their roles are given in Table 1. Two of these resources (Adelaide and Sydney) were effectively functioning as single processor machines as the Symmetric Multi-Processing (SMP) Linux kernel was not running on them. All the machines in this testbed were running Globus 2.4.2 and NWS sensors. The broker was deployed on the Melbourne Computer Science machine and broker agents were dispatched at runtime to the other resources for executing jobs and initiating data transfers.

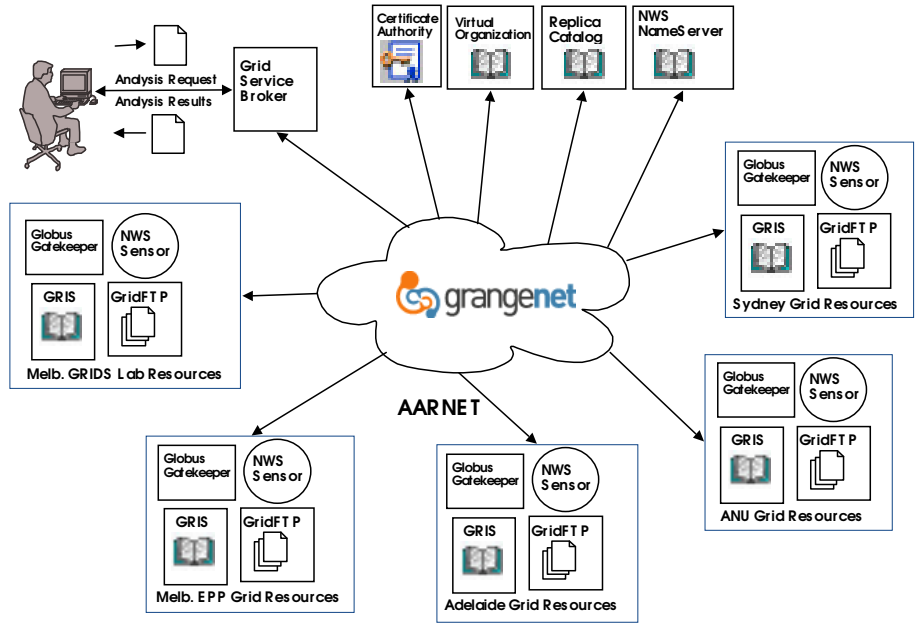

**Figure 5: Australian Belle Analysis Data Grid testbed.**

Data that was produced on one site in BADG had to be shared with the other sites. For this purpose, a Data Catalog was set up for the Belle Data Grid using the Globus Replica Catalog (RC) mechanism. The Data Catalog is a global directory structure that stores the logical file names and their physical locations and is based on the Lightweight Directory Access Protocol (LDAP). A set of high-level tools emulating Unix directory structure commands for creation and management of RC have also been developed by utilitsing low-level Globus RC functions.

The primary application for the Belle experiment is the Belle Analysis Software Framework (BASF). This application is used for simulation, filtering of events, and analysis. It is also a legacy application that consists of about a Gigabyte of code. Therefore, execution is restricted to those sites which have this application already installed. In case the data is being transferred across the network, a certain amount of time results before the data is completely available on the executing node. To eliminate this "dead-time", BASF was modified to execute on streaming data.

### 4.3  Application Parameterisation and Experimental Setup

A typical analysis is split into two streams: data and simulation. Raw data is recorded from various sensors within a detector and stored as separate measurements or "events". Simulated or Monte-Carlo data involves the generation of events and then detailed detector simulation. From this point on, the analysis streams are very similar. The data is reconstructed, which involves the correlation of sensor information. Data summaries are generated for ease of analysis. As an example, within the Belle experiment 10 TB of data summary information exists. These are "skimmed" to produce subsets of the data of most interest to each physicist's analysis. These are around 100 GB in size for Belle users. These are then analysed to generate plots and histograms and can then be used for statistical analysis by applying further cuts. For simulated data, this process is repeated until the analysis is perfected. The simulated data can then be used for systematic error analysis. The same analysis process is performed on data to obtain a result, provided there are no large differences between data and simulation. Simulation is very CPU-intensive and the results must be saved and made available to the whole collaboration.

For validating the broker, a simulation of a "decay chain" of particles has been used. A decay chain occurs when an unstable particle decays into another and so on until a stable particle state is reached. This is typical of the events that happen within a particle accelerator. The experiment consists of 2 parts, both of which involve execution over the Grid using the Gridbus broker. In the first part, 100,000 events of the decay chain $B^0 \rightarrow D^{*+}D^{*-}K_s$ shown in Figure 6 are simulated via distributed generation and this data is entered into the replica catalog. In the analysis part, the replica catalog is queried for the generated data and this is analysed over the Belle Data Grid. The histograms resulting from this analysis are then returned as output. Here only the results of the analysis are discussed as it involved accessing remote data.

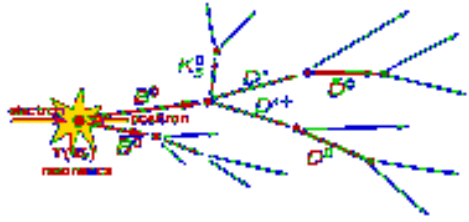

**Figure 6: The $B^0$->D*+D*-Ks decay chain.**

A plan file for the composing analysis of Belle data as a parameter sweep application is shown in Figure 7. The plan file follows Nimrod-G's declarative parametric programming language which has been extended in this work by introducing a new type of parameter called "Gridfile". This dynamic parameter describes a logical file location, either a directory or a collection of files and the broker resolves it to the actual file names and their physical locations. The plan file also instructs copying of user defined analysis modules and configuration files to the remote sites before any execution is started. The main task involves executing a user-defined shell script at the remote site which has 2 input parameters: the full network path to the data file and the name of the job itself. The shell script invokes BASF at the remote site to conduct the analysis over the data file and produce histograms. The histograms are then copied over to the broker host machine.

**Table 1: List of resources used in the experiment**

| Organization | Machine details | Role |
|---|---|---|
| Dept. of Computer Science, University of Melbourne. | `belle.cs.mu.oz.au` IBM eServer, 4 CPU, 2GB RAM, 70 GB HD, RH Linux 8.0 | Broker host, Data host, Compute resource, NWS server |
| School of Physics, University of Melbourne | `fleagle.ph.unimelb.edu.au` PC, 1 CPU, 512 MB RAM, 70 GB HD, RH Linux 8.0 | Replica Catalog host, VO Tools host, Data host, Compute resource, NWS sensor |
| Dept. of Computer Science, University of Adelaide | `belle.cs.adelaide.edu.au` IBM eServer, 4 CPU (only 1 available), 2GB RAM, 70 GB HD, RH Linux 7.3 | Data host, Compute resource, NWS sensor |
| Australian National University, Canberra | `belle.anu.edu.au` IBM eServer, 4 CPU, 2GB RAM, 70 GB HD, RH Linux 7.3 | Data host, Compute resource, NWS sensor |
| Dept of Physics, University of Sydney | `belle.physics.usyd.edu.au` IBM eServer, 4 CPU (only 1 available), 2GB RAM, 70 GB HD, RH Linux 7.3 | Data host, Compute resource, NWS sensor |

The Logical file name in this particular experiment resolved to 100 Monte Carlo simulation data files. Therefore, the experiment set consisted of 100 jobs, each dealing with the analysis of one data file using BASF. Each of these input data files was 30 MB in size. The entire data set was equally distributed among the five data hosts i.e. each of them has 20 data files each. The data was also not replicated between the resources, therefore, the dataset on each resource remained unique to it. The histograms generated were 968 KB in size and online visualization of histogram outputs is shown in Figure 8.

For monitoring the bandwidth between the resources, an NWS sensor was started on each of the resources which reports to the NWS name server located in Melbourne. An NWS activity for monitoring bandwidth was defined at the name server within which a clique containing all the resources on the testbed was created. Members of the clique conduct experiments one at a time to determine network conditions between them. Querying the name server at any point provides the bandwidth and latency between any 2 members of the clique.

**New parameter type defined to describe an input data file**

**Logical file name pointing to the location in the replica catalog**

```
parameter INFILE Gridfile lfn:/users/winton/fsimddks/fsimdata*.mdst;
task nodestart
     copy ddks_ana.so node:ddks_ana.so
     copy libanalyser.so node:libanalyser.so
     copy libbase_analyser.so node:libbase_analyser.so
     copy libreconstructor.so node:libreconstructor.so
     copy libtools.so node:libtools.so
     copy event.conf node:event.conf
     copy recon.conf node:recon.conf
     copy particle.conf node:particle.conf
endtask
task main
     node:execute ./runme.ddksana $INFILE $jobname
     copy node:runme.log runme.log.$jobname
     copy node:ddks-$jobname.hbook ddks-$jobname.hbook
endtask
```

**Figure 7: Plan file for Data Analysis.**

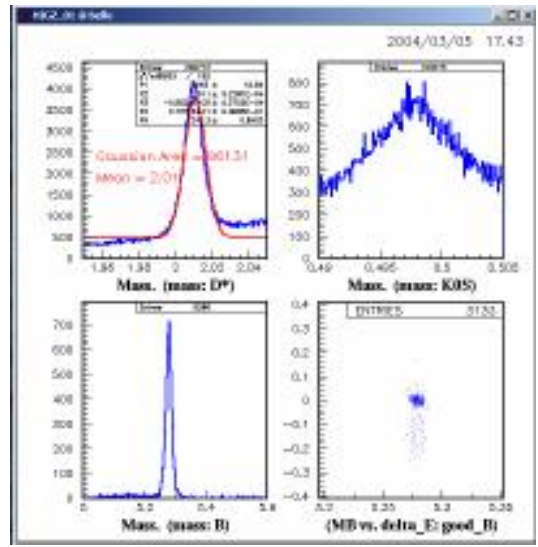

**Figure 8: A histogram generated from Belle analysis.**

### 4.4 Results of Evaluation

Three scheduling scenarios were evaluated: (1) scheduling with computation limited to only those resources with data, (2) scheduling without considering location of data, and (3) our adaptive scheduling (shown in Figure 3) that optimizes computation based on the location of data. The experiments were carried out on April 19[th], 2004 between 18:00 and 23:00 AEST. At that time, the Globus gatekeeper service on the Adelaide machine was down and so, it could not be used as a computational resource. However, it was possible to obtain data from it through GridFTP. Hence, jobs that depended on data hosted on the Adelaide server were able to be executed on other machines in the second and third strategies. A graph depicting the comparison of the total time taken for each strategy to execute all the jobs is shown in Figure 9 and another comparing resource performance for different scheduling strategies is shown in Figure 10

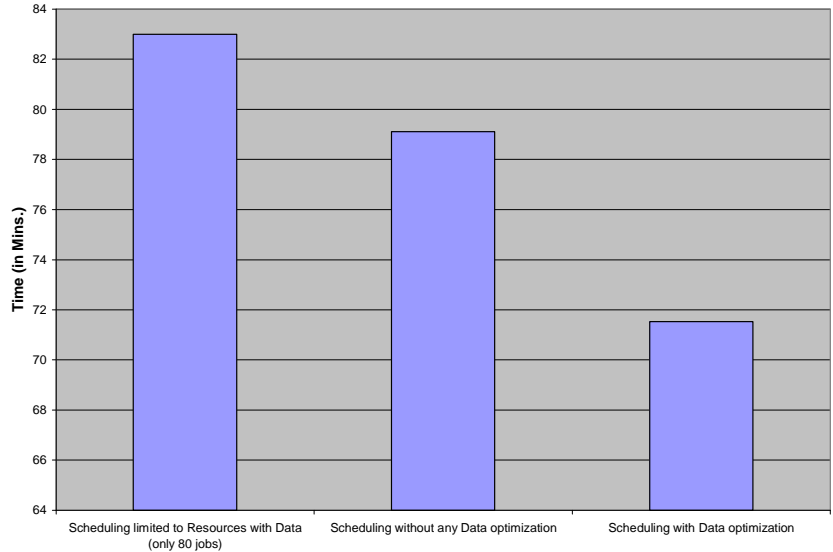

**Figure 9: Total time taken for each scheduling strategy.**

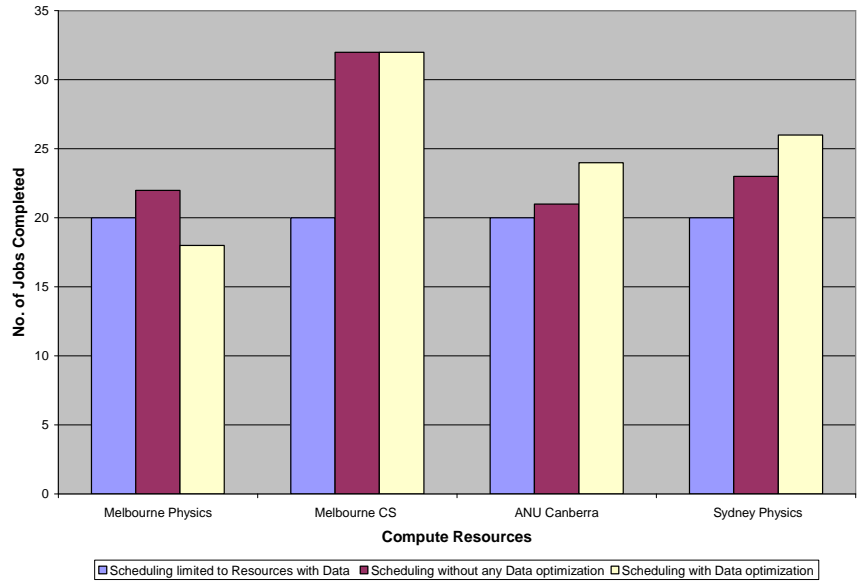

**Figure 10: Comparison of resource performance under different scheduling strategies.**

In the first strategy (scheduling limited to resources with the data for the job), jobs were executed *only* on those resources which hosted the data files related to those jobs. No data transfers were involved in this scenario. As is displayed in the graph in Figure 10, all of the resources except the one in Adelaide were able to execute 20 jobs each. The jobs that were scheduled on that resource failed, as its computational service was unavailable. Hence, Figure 9 shows the total time taken for only 80 successful jobs out of 100. However, this time also includes the time taken by the scheduler to analyse the remaining 20 jobs as failed. In this setup, the related data was *exclusively* located on that resource and hence, these jobs were not reassigned to other compute resources. Thus, a major disadvantage of this scheduling strategy was exposed.

In the second strategy (scheduling without any data optimization), the jobs were executed on those nodes that have the most available computational resources. That is, there was no optimization based on location of data within this policy. The Adelaide server was considered a failed resource and was not given any jobs. However, the jobs that utilized data files hosted on this machine were able to be executed on other resources. This strategy involves the maximum amount of data transfer which makes it unsuitable for applications involving large data transfers and utilising resources connected by slow networks.

The last evaluation (scheduling with data optimization) was carried out by scheduling jobs to the compute resources that satisfied the algorithm given in Section 3.3. In this case, as there were no multiple data hosts for the same data, the policy was reduced to dispatching jobs to the best available compute resource that had the best available bandwidth to the host for the related data. It can be seen from Figure 10 that most of the jobs that accessed data present on the Adelaide resource were scheduled on the Melbourne Physics and CS resources because the latter had consistently higher available bandwidth to the former. This is shown in the plot of the available bandwidth from the University of Adelaide to other resources within the testbed measured during the execution, given in Figure 11. The NWS name server was polled every scheduling interval for the bandwidth measurements. As can be seen from Figure 9, this strategy took the least time of all three.

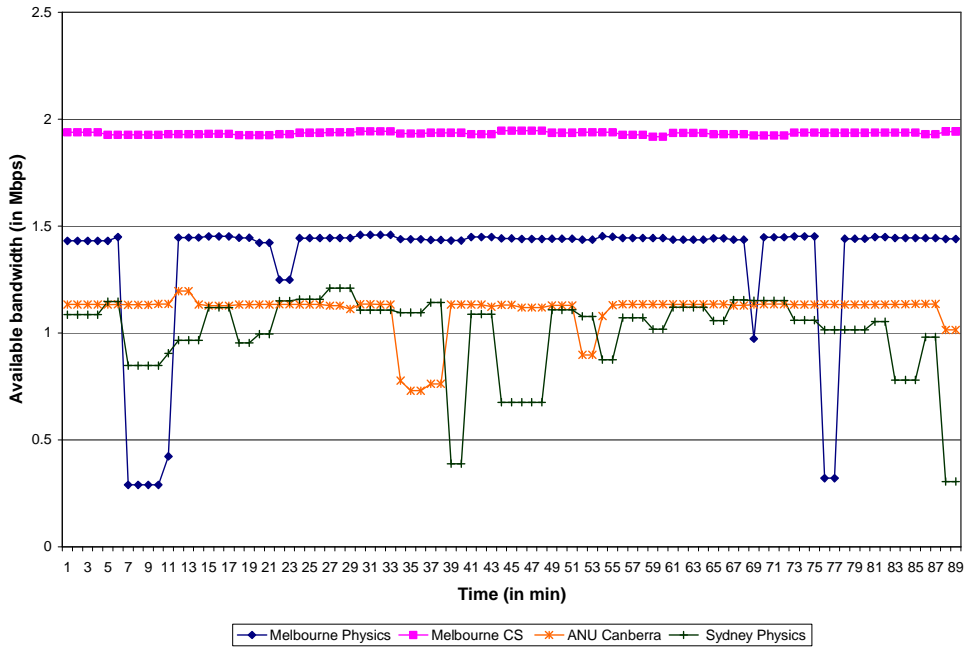

**Figure 11: Available bandwidth from University of Adelaide to other resources in the testbed.**

## 5  Summary and Conclusion

We have presented a grid broker for executing distributed data-oriented jobs on a grid. The broker discovers computational and data resources, schedules jobs based on optimization of data transfer and returns results back to the user.  We have applied this broker to a data-intensive environment, which is the analysis of the Belle high-energy physics experiment data and have presented the results of our evaluation with different scheduling strategies. The scheduling strategy proposed within the broker took into consideration the network conditions and has produced the best possible outcome by executing the jobs within the least amount of time.

We plan to conduct further evaluations with larger file sizes and multiple repositories for the same datasets. This will ensure that the data transfer time becomes more significant while making scheduling decisions and that the scheduler will be able to choose between different data hosts. We plan to extend the broker by completely integrating economic mechanisms described in Section 3.1 within the architecture.

## Acknowledgement

The authors would like to express their gratitude to members of the Gridbus Project and collaborators: Glenn Moloney and Martin Sevior (School of Physics) for sharing their thoughts on grid-enabling the Belle analysis framework, Jia Yu for her contribution towards the implementation of job wrapper for Globus, Hussein Gibbins and Shoaib Burq for their contribution towards plan file interpretation, Akshay Luther for his design and implementation of Alchemi, Rajiv Ranjan for his contribution towards implementation of adapter for Alchemi and Nimrod-G protocol interface, Ding Choon Hong for his contribution towards integration of NWS within the broker (GRIDS Lab), Brett Beeson for extending the copy command to support wildcards and using the broker services to develop a portal for Astrophysics, Steve Melnikoff for his comments and using the broker services to develop a portal for Belle analysis (School of Physics), Huy Le (University of Adelaide) for exploring the use of Gridbus broker in data-aware resource scheduling and Benjamin Khoo (IBM, Singapore) for his technical inputs during implementation. We gratefully acknowledge IBM for their donation of 4 eServer machines to the Belle testbed and VPAC for providing access to their cluster and the system coordinators of the testbed resources located in Adelaide University (Andrew Wendelborn and Paul Coddington), Australian National University (Markus Buchorn) and Sydney University (Kevin Varvell).

## References

[1] I. Foster and C. Kesselman (editors), *The Grid: Blueprint for a Future Computing Infrastructure*, Morgan Kaufmann Publishers, USA, 1999.

[2] A. Chervenak, I. Foster, C. Kesselman, C. Salisbury, and S. Tuecke, The data grid: Towards an architecture for the distributed management and analysis of large scientific datasets*, Journal of Network and Computer Applications*, vol. 23, no. 3, pp. 187–200, 2000.

[3] R. Moore, C. Baru, R. Marciano, A. Rajasekar, and M.Wan, *The Grid: Blueprint for a New Computing Infrastructure.* Morgan Kaufmann, 1998, ch. 5, "Data Intensive Computing", pp. 105–131.

[4] C. Baru, R. Moore, A. Rajasekar, and M. Wan, *The SDSC Storage Resource Broker*, in CASCON'98 Conference, Toronto, Canada, 1998.

[5] Asad Samar, Heinz Stockinger. *Grid Data Management Pilot (GDMP): A Tool for Wide Area Replication*, IASTED International Conference on Applied Informatics (AI2001)*,* Innsbruck, Austria, February 2001.

[6] A. Chervenak, E. Deelman, I. Foster, L. Guy, W. Hoschek, A. Iamnitchi, C. Kesselman, P. Kunst, M. Ripeanu, B, Schwartzkopf, H, Stockinger, K. Stockinger, B. Tierney. *Giggle: A Framework for Constructing Scalable Replica Location Services*, Proceedings of Supercomputing 2002 (SC2002), November 2002.

[7] G. Allen, W. Benger, T. Goodale, H. Hege, G. Lanfermann, A. Merzky, T. Radke, E. Seidel, J. Shalf, *The Cactus Code: A Problem Solving Environment for the Grid*, Proceedings of the Ninth International Symposium on High Performance Distributed Computing (HPDC), Pittsburgh, USA, IEEE Press.

[8] W. Hoschek, J. Jaen-Martinez, A. Samar, H. Stockinger, K. Stockinger, *Data Management in an International Data Grid Project*, Proceedings of the 1st International Workshop on Grid Computing (Grid 2000, Bangalore, India), Springer-Verlag, Berlin, Germany, 2000.

[9] H. Casanova, G. Obertelli, F. Berman, and R. Wolski, *The AppLeS Parameter Sweep Template: User-Level Middleware for the Grid*, Proceedings of the IEEE SC 2000: International Conference Networking and Computing, Nov. 2000, Dallas, Texas, IEEE CS Press, USA.

[10] D. Abramson, J. Giddy, and L. Kotler, *High Performance Parametric Modeling with Nimrod/G: Killer Application for the Global Grid?*, Proceedings of the International Parallel and Distributed Processing Symposium (IPDPS 2000), May 1-5, 2000, Cancun, Mexico, IEEE CS Press, USA, 2000.

[11] R. Buyya, D. Abramson, and J. Giddy, *An Economy Driven Resource Management Architecture for Global Computational Power Grids*, Proceedings of the 2000 International Conference on Parallel and Distributed Processing Techniques and Applications (PDPTA 2000), June 26-29, 2000, Las Vegas, USA, CSREA Press, USA, 2000.

[12] K. Ranganathan and I. Foster, *Decoupling Computation and Data Scheduling in Distributed Data-Intensive Applications*, Proceedings of 11th IEEE International Symposium on High Performance Distributed Computing (HPDC-11), Edinburgh, Scotland, July 2002, IEEE CS Press, USA,.

[13] *The ATLAS Experiment, CERN.* http://atlas.web.cern.ch/Atlas/Welcome.html (Accessed Jan 2004).

[14] *The CMS Experiment, CERN.* http://cmsinfo.cern.ch/Welcome.html (Accessed Jan 2004).

[15] *The UK eScience Programme.* http://www.rcuk.ac.uk/escience/ (Accessed Feb 2004).

[16] *The Belle experiment, KEK.* http://belle.kek.jp/ (Accessed Jan 2004).

[17] I. Foster, C. Kesselman, and S. Tuecke, The anatomy of the grid: Enabling scalable virtual organizations, *International Journal of High Performance Computing Applications*, vol. 15, pp. 200-222, Sage Publishers, London, UK, 2001.

[18] S. Vazhkudai, S. Tuecke, I. Foster, *Replica Selection in the Globus Data Grid*, Proceedings of the First IEEE/ACM International Conference on Cluster Computing and the Grid (CCGRID 2001), pp. 106-113, IEEE Computer Society Press, May 2001.

[19] *The MONARC Project, CERN*, http://monarc.web.cern.ch/MONARC/ (Accessed Jan 2004).

[20] *The Large Hadron Collider, CERN,* http://lhc-new-homepage.web.cern.ch/lhc-new-homepage/ (Accessed Jan 2004).

[21] J. Yu and R. Buyya, *Grid Market Directory: A Web and Web Services based Grid Service Publication Directory*, Technical Report, GRIDS-TR-2003-0, Grid Computing and Distributed Systems (GRIDS) Laboratory, The University of Melbourne, Australia, January 2003.

[22] R. Buyya, J. Giddy, and D. Abramson, *A Case for Economy Grid Architecture for Service-Oriented Grid Computing*, 10th IEEE International Heterogeneous Computing Workshop (HCW 2001), In conjunction with IPDPS 2001, San Francisco, California, USA, April 2001

[23] A. Barmouta and R. Buyya, *GridBank: A Grid Accounting Services Architecture (GASA) for Distributed Systems Sharing and Integration*, Workshop on Internet Computing and E-Commerce, Proceedings of the 17th Annual International Parallel and Distributed Processing Symposium (IPDPS 2003), IEEE Computer Society Press, USA, April 22-26, 2003, Nice, France.

[24] K. Holtman et al, *CMS Requirements for the Grid*, Proc. of CHEP 2001 (Beijing, September 3 - 7, 2001)*,* p. 754-757. Science Press. ISBN 1-880132-77-X

[25] K. Czajkowski, S. Fitzgerald, I. Foster, and C. Kesselman, *Grid Information Services for Distributed Resource Sharing*, Proceedings of 10th IEEE International Symposium on High Performance Distributed Computing (HPDC-10), IEEE CS Press, USA, 2001.

[26] J. Bester, I. Foster, C. Kesselman, J. Tedesco, S. Tuecke, *GASS: A Data Movement and Access Service for Wide Area Computing Systems,* Proceedings of the Sixth Workshop on Input/Output in Parallel and Distributed Systems, pages 78-88, Atlanta, GA, May 1999. ACM Press.

[27] G. von Laszewski, I. Foster, J. Gawor, and P. Lane, A Java Commodity Grid Kit, *Concurrency and Computation: Practice and Experience*, vol. 13, no. 8-9, pp. 643-662, 2001, http:/www.cogkits.org/.

[28] J. Bunn and H. Newman, Data-intensive grids for high-energy physics, In *Grid Computing: Making the Global Infrastructure a Reality*, F. Berman, G. Fox, and T. Hey, Eds. John Wiley & Sons, Inc., New York, 2003.

[29] Lyle Winton, Data Grids and High Energy Physics: A Melbourne Perspective, *Space Science Reviews*, 107 (1-2): 523-540, Kluwer Academic Publishers, Netherlands, 2003

[30] K. Marzullo, M. Ogg, A. Ricciardi, A. Amoroso, F. Calkins, E. Rothfus, *NILE: Wide-Area Computing for High Energy Physics*, Proceedings of 7th ACM SIGOPS European Workshop, Connemara, Ireland, 2-4 Sept. 1996, ACM Press.

[31] J.H. Christenson, et al *Phys. Rev. Lett*. 13 (1964) 138-140.

[32] *Australian Belle Analysis Data Grid,* http://roberts.ph.unimelb.edu.au/epp/grid/badg/. Accessed Feb 2004

[33] R. Wolski, N. Spring, and J. Hayes, "The Network Weather Service: A Distributed Resource Performance Forecasting Service for Metacomputing", *Journal of Future Generation Computing Systems*,Volume 15, Numbers 5-6, pp. 757-768, Elsevier Science.

[34] Ian Foster and Carl Kesselman, "Globus: A Metacomputing Infrastructure Toolkit", *International Journal of Supercomputer Applications*, 11(2): 115-128, 1997.

[35] GrangeNet (GRid And Next GEneration Network), http://www.grangenet.net, Accessed Feb. 2004.

[36] A. Luther, R. Buyya, R. Ranjan, and S. Venugopal, *Alchemi: A .NET-based Grid Computing Framework and its Integration into Global Grids*, Technical Report, GRIDS-TR-2003-8*,* Grid Computing and Distributed Systems Laboratory, University of Melbourne, Australia, December 2003.